\documentclass[10pt,letterpaper]{article}
\usepackage[top=0.85in,footskip=0.75in]{geometry}
\usepackage{float}
\usepackage{amsmath,amssymb}

\usepackage{listings}

\usepackage{changepage}

\usepackage{textcomp,marvosym}

\usepackage{cite}

\usepackage{nameref,hyperref}

\usepackage[nopatch=eqnum]{microtype}
\DisableLigatures[f]{encoding = *, family = * }

\usepackage[table]{xcolor}

\usepackage{array}

\newcolumntype{+}{!{\vrule width 2pt}}

\newlength\savedwidth

\raggedright
\usepackage[aboveskip=1pt,labelfont=bf,labelsep=period,justification=raggedright,singlelinecheck=off]{caption}

\makeatletter
\renewcommand{\@biblabel}[1]{\quad#1.}
\makeatother

\usepackage{lastpage,fancyhdr,graphicx}
\usepackage{epstopdf}
\fancyheadoffset[L]{2.25in}
\fancyfootoffset[L]{2.25in}
\begin{document}
\vspace*{0.2in}

% Title must be 250 characters or less.
\begin{flushleft}
{\Large
\textbf\newline{Can ephapticity contribute to the brain complexity?} % Please use "sentence case" for title and headings (capitalize only the first word in a title (or heading), the first word in a subtitle (or subheading), and any proper nouns).
}
\newline
% Insert author names, affiliations and corresponding author email (do not include titles, positions, or degrees).
\\
Gabriel Moreno Cunha\textsuperscript{1,4\P},
Gilberto Corso\textsuperscript{1,2\&},
Matheus Phellipe Brasil de Sousa\textsuperscript{1,4\&},
Gustavo Zampier dos Santos Lima\textsuperscript{1,3,4,5*\&}
%\textsuperscript{\textpilcrow}
\\
\bigskip
\textbf{1} Departamento de Física Teórica e Experimental, Universidade Federal do Rio Grande do Norte, 59078-970, Natal, RN, Brazil.
\\
\textbf{2} Departamento de Biofísica e Farmacologia, Universidade Federal do Rio Grande do Norte, 59078-970, Natal, RN, Brazil.
\\
\textbf{3} Escola de Ciências e Tecnologia, Universidade Federal do Rio Grande do Norte, 59078-970, Natal, RN, Brazil.
\\
\textbf{4} Laboratório de Simulação e Modelagem Neurodinâmica, Universidade Federal do Rio Grande do Norte, 59078-970, Natal, RN, Brazil.
\\
\textbf{5} Institut Camille Jordan, UMR 5208 CNRS, University Lyon 1, 69622 Villeurbanne, France
\bigskip

% Insert additional author notes using the symbols described below. Insert symbol callouts after author names as necessary.
% 
% Remove or comment out the author notes below if they aren't used.
%
% Primary Equal Contribution Note
\P These authors contributed equally to this work.\\
\& These authors also contributed equally to this work.

% Additional Equal Contribution Note
% Also use this double-dagger symbol for special authorship notes, such as senior authorship.
%\ddag These authors also contributed equally to this work.

% Current address notes
%\textcurrency Current Address: Dept/Program/Center, Institution Name, City, State, Country % change symbol to "\textcurrency a" if more than one current address note
% \textcurrency b Insert second current address 
% \textcurrency c Insert third current address

% Deceased author note
%\dag Deceased

% Group/Consortium Author Note
%\textpilcrow Membership list can be found in the Acknowledgments section.

% Use the asterisk to denote corresponding authorship and provide email address in note below.
* gustavo.zampier@ufrn.br

\end{flushleft}
% Please keep the abstract below 300 words
\section*{Abstract}

The inquiry into the origin of brain complexity remains a pivotal question in neuroscience. While synaptic stimuli are acknowledged as significant, their efficacy often falls short in elucidating the extensive interconnections of the brain and nuanced levels of cognitive integration. Recent advances in neuroscience have brought the mechanisms underlying the generation of highly intricate dynamics, emergent patterns, and sophisticated oscillatory signals into question. Within this context, our study, in alignment with current research, posits the hypothesis that ephaptic communication may emerge as the primary candidate for unraveling optimal brain complexity. In this investigation, we conducted a comparative analysis between two types of networks utilizing the Quadratic Integrate-and-Fire Ephaptic model (QIF-E): (I) a small-world synaptic network (ephaptic-off) and (II) a mixed composite network comprising a small-world synaptic network with the addition of an ephaptic network (ephaptic-on). Utilizing the Multiscale Entropy methodology, we conducted an in-depth analysis of the responses generated by both network configurations, with complexity assessed by integrating across all temporal scales. Our findings demonstrate that ephaptic coupling enhances complexity under specific topological conditions, considering variables such as time, spatial scales, and synaptic intensity. These results offer fresh insights into the dynamics of communication within the nervous system and underscore the fundamental role of ephapticity in regulating complex brain functions. 
% Please keep the Author Summary between 150 and 200 words
% Use first person. PLOS ONE authors please skip this step. 
% Author Summary not valid for PLOS ONE submissions.   
%\linenumbers

% Use "Eq" instead of "Equation" for equation citations.
\section*{Introduction}

The brain can be understood as a sophisticated system in which mental states arise from interactions that span multiple levels encompassing physical and functional aspects \cite{nunez2012brain, bullmore2009generic, bassett2011understanding}. The human mind is an intricate phenomenon that develops beneath the structural complexity of the brain \cite{hagmann2008mapping, bassett2011conserved}. However, the precise nature of the mind-brain connection remains elusive, and a full understanding has yet to be achieved. The structure of the brain spans multiple temporal and spatial dimensions, giving rise to sophisticated cellular and neuronal phenomena that collectively constitute the physical basis of cognition \cite{he2010temporal}. In the spatial dimension, the cerebral organization  exhibits similar patterns at various resolutions in the distribution of cells throughout the brain \cite{bullmore2012economy, moser2008place}. In the temporal dimension, for example, there are modules for short-term and long-term memory \cite{shallice1996domain,lima2014mouse}. The architecture of the brain is inextricably linked to its connectivity, both in terms of function and structure \cite{petersen2015brain}.

The brain can be viewed as a highly complex system where mental states emerge from interactions spanning multiple levels, encompassing both physical and functional aspects \cite{nunez2012brain, bullmore2009generic, bassett2011understanding}. The human mind is understood as a structure of intricate complexity, given by the vast interconnected neural network of the brain. This phenomenon encompasses all of human consciousness, cognition, emotions and perceptions, manifesting as a symphony of interconnected processes that shape our understanding of the world \cite{hagmann2008mapping, bassett2011conserved}. Despite significant advancements, the precise nature of the mind-brain connection remains elusive, with a comprehensive understanding yet to be attained. Spanning multiple temporal and spatial dimensions, the brain's structure gives rise to sophisticated cellular and neuronal phenomena that collectively form the physical basis of cognition \cite{he2010temporal}. Spatially, the brain's organization exhibits similar patterns across various resolutions in the distribution of cells throughout its regions \cite{bullmore2012economy, moser2008place}. Temporally, the brain comprises modules dedicated to short-term and long-term memory, among other functions \cite{shallice1996domain, lima2014mouse}. The brain's architecture is intricately intertwined with its connectivity, both functionally and structurally \cite{petersen2015brain}.

The flow of information between neurons through synaptic firing patterns has always been considered the fundamental basis of neuronal processes, encompassing essential functions such as memory and consciousness \cite{hobson2002cognitive, shepherd2003synaptic,queenan2017research}. As we advance in our understanding of the brain, it is increasingly recognized that the complexity of neuronal communication goes beyond synaptic connectivity \cite{queenan2017research}. In addition to synapses, adjacent electrical fields, known as ephaptics, are emerging as protagonists in modulating neuronal architecture and influencing functional responses \cite{hagmann2008mapping, bassett2011conserved,su2012non,zhang2019asymmetric}. Ephaptic communication refers to cases in which neighboring neurons establish electrical connections and modulate extracellular flow \cite{anastassiou2011ephaptic,anastassiou2010effect,shifman2019elfenn}. This subtle electrical interaction highlights the harmonious interconnection that goes beyond synapses and adds a new dimension to the understanding of communication in the brain. Ephapticity emerges as narrative of neuronal complexity, suggesting that brain communication may transcend the boundaries of known synapses.\cite{shifman2019elfenn, jefferys1995nonsynaptic}. Due to the short range of the electric fields, the ephapticity generated by a neuron affects neighboring neurons \cite{su2012non,zhang2019asymmetric}. This phenomenon has also been observed in a study in which electrical inhibition was induced in rat cells \cite{weiss2008role}. Furthermore, ephaptic coupling has been identified as a crucial factor in governing synchronization and spike timing in neurons \cite{han2018ephaptic, schmidt2021ephaptic, anastassiou2011}.

In the study, coworker presented a compelling argument suggesting that memory formation in the brain is associated with ephaptic processes that intrinsically shape and control neuronal activity by establishing connections between the brain areas \cite{pinotsis2023vivo}. Their study provided empirical evidence for ephaptic coupling between two cortical regions in vivo. The results strongly suggest that ephaptic coupling, driven by electric fields, plays a causal role in local neuronal activity. It is interesting to emphasize that neuronal activity under the influence of ephaptic coupling transmitted less information and exhibited greater variability and complexity. In another study, Hunt et al. provided convincing evidence that oscillating electromagnetic (EM) fields play a pivotal role in steering and unifying conscious cognition \cite{hunt2023fields}. Their study suggests that EM fields are not just by-products of brain functions but that they trigger various crucial functions. There is a possibility that the brain's local and global electromagnetic fields may actually serve as a central locus of consciousness \cite{hunt2023fields}.

Neuronal ephaptic communication, which is essential for neuronal function, increases complexity through direct electrical interactions. This phenomenon, often overlooked in neuronal models, emphasizes the need for a more comprehensive understanding of the intricate processes in the brain. With this in mind, in this work we explore the role of ephaptic communication in parallel with synaptic networks, using a small-world network to model of the brain complexity \cite{masuda2004global,liu2022analysis,bassett2011understanding,guardiola2000synchronization,sporns2022structure,bassett2006small}. To this end, we performed various simulations of network structures using a small-world synaptic topology in two cases: (I) synaptic network only (ephaptic-off); (II) synaptic and ephaptic network (ephaptic-on). For the ephaptic-off network, we changed the following topological parameters: the number of neurons (\textit{N}), the rewiring probability (\textit{rp}) of the small world, and the synaptic intensity ($\omega^{(k)}$) of the connection. Moreover, the coupling of the ephaptic network is all-to-all with weights that depend on the distance of  the neurons. To quantify the complexity, we used the multiscale entropy integration, a Shannon-like entropy developed for several time scales. The multiscale entropy (MSE) is based on the work of Zhang and the method of Costa et al. \cite{costa2002multiscale, costa2008multiscale, zhang1991complexity}. Our results show not only that ephaptic communication necessarily contributes to explaining the complexity of the brain, but also that the balance between synaptic and ephaptic processes is essential for maintaining brain functionality.

This paper is divided into four sections as follows: The Methods section shows how the QIF-E model is developed by the current ephaptic coupling approach. We then present the network model, both for small-world ephaptic-off networks and for networks with ephaptic coupling. We also discuss the mathematical tool called Multiscale Entropy (MSE), which quantifies the complexity of the neuronal network. In the Results section, the numerical simulations and data analysis are presented. Finally, the Discussion section provides a new perspective for understanding the complexity of the brain by considering the balance between synaptic and ephaptic communication.

\section*{Materials and methods}
\subsection*{Firing neuron model with Ephaptic Coupling}
\noindent
The quadratic integrate- and-fire model with ephapticity (QIF-E) \cite{cunha2022ephaptic, cunha2023electrophysiological} is a simplified neuron model, it is an  integrate- and-fire neuron model that describes spikes in neurons inserted in an electric field given by the LFP, referred to here as the ephaptic term (equation (\ref{eq:qif_adapt})). In contrast to physiologically accurate but computationally expensive neuronal models, the QIF-E model generates a standard action potential-like pattern and ignores subtleties such as control variables. According to Cunha et al. \cite{cunha2022ephaptic} ephaptic communication can be simulated by the following QIF-E hybrid model:
\begin{equation}
    \Dot{V}_{m}(t) \quad =  \quad \underbrace{a . V^{2}_{m}(t)+b . V_{m}(t)}_{\text{\bf{Cell intrinsic dynamic term}}}  \quad -  \quad \underbrace{ c .I_{ephap}(t)}_{\text{\bf{Ephaptic term}}}  \quad +  \quad\underbrace{ I_{0}(t)}_{\text{\bf{Synaptic term}}} ,
    \label{eq:qif_adapt}
\end{equation}

\noindent
where $V_{m}$ is the membrane potential difference, $a$ and $b$ are parameters related to the electrical properties of the neuron membrane, such as membrane resistance and capacitance. The parameter $c$ is the ephaptic weight, which is based on the electrophysiological properties of the extracellular and membrane milieu. Finally, the current terms ($I_{ephap}$ and $I_{0}$) are related to communication stimuli.

\begin{figure}[H]
    \centering
    \includegraphics[width=1\textwidth]{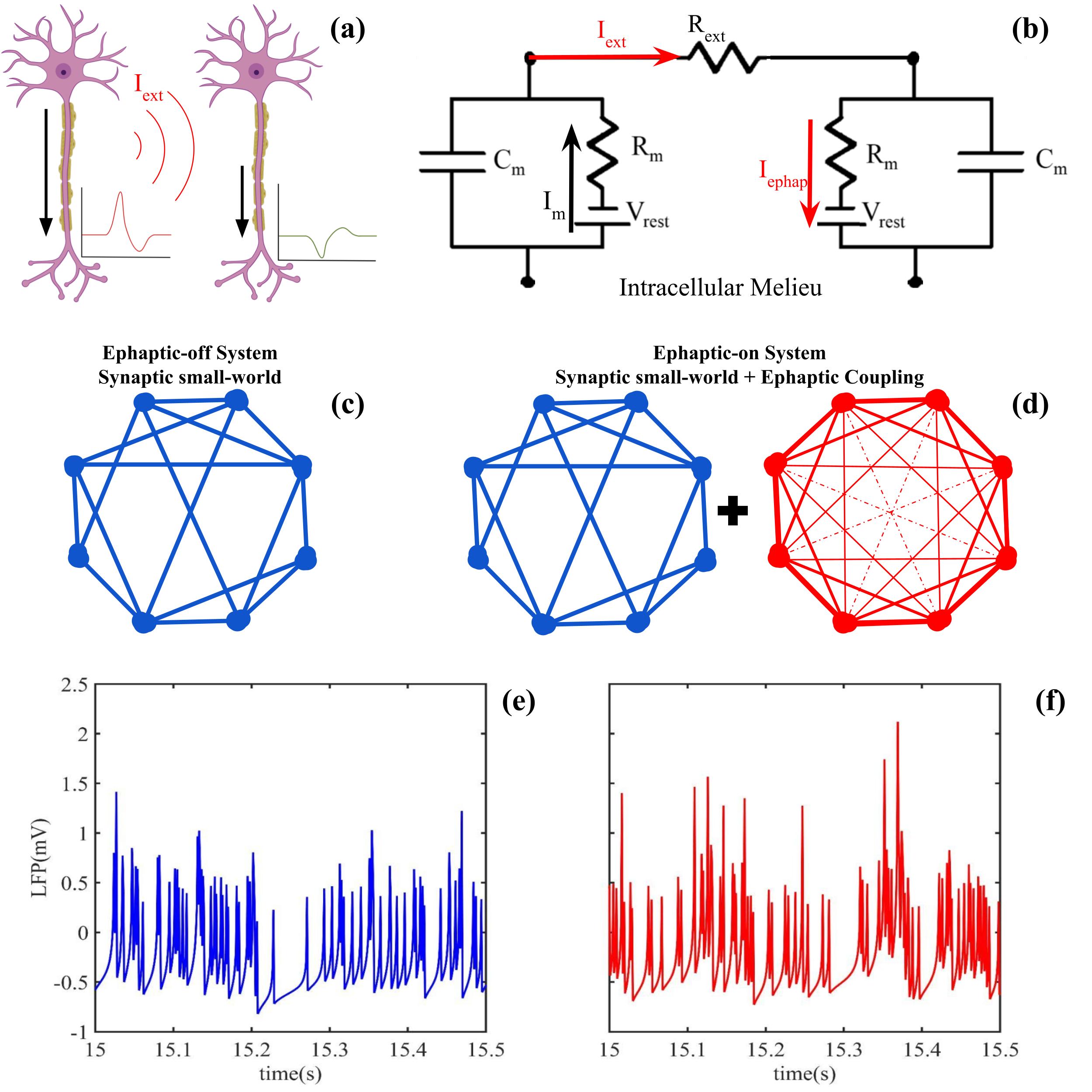}
    \caption{\textbf{Ephaptic Coupling Model.} (a) Representation of ephaptic coupling mechanism. (b) Equivalent circuit to ephaptic coupling in QIF-E model (c) The ephaptic-off network is modeled by small-world topology. (d) The brain network is mimetized  by a ephaptic-on network, which is distance dependent. The distance between neurons are represented by red line width. (e) Ephaptic-off tempoal serie provided by the mean of all neurons in the network. (f) Ephaptic-on tempoal serie provided by the mean of all neurons in the network.}
    \label{fig:fig1}
\end{figure}

To perform the ephaptic coupling in the QIF-E, we assume that all neurons are approximately spherical (SOMA), and we do not consider the propagation effects of the spikes along the axon. Therefore, the ephaptic term in equation (\ref{eq:qif_adapt}) is estimated assuming that the membrane behaves like an electrical circuit (see Fig. \ref{fig:fig1} (b)). Assuming that the membrane of neurons $i$ and $j$ can be modeled by this description, the ephaptic current in equation (\ref{eq:qif_adapt}) is given by:

\begin{equation}
    c.I_{ephap}(t) = \frac{V_{ephap}(t)}{C_{m}R_{ext}} 
    \label{eq:ephap}
\end{equation}

\noindent
where $C_{m}$ is the membrane capacitance, $V_{m}$ is the membrane potential difference, $a$ and $b$ are parameters related to the electrical properties of the neuron membrane, and current terms are related to the communication stimuli, $R_{ext}$ is the extracellular resistance. In the approach by Shifman \& Lewis\cite{shifman2019elfenn}, which uses the concepts of other works\cite{holt&koch1999, logothetis2007vivo, anastassiou2011,mechler2012dipole,cunha2022ephaptic}, the equation (\ref{eq:ephap}) can be described by:

\begin{equation}
    c.I_{ephap}^{(1)}(t) = \frac{V_{m}^{(1)}(t)-V_{m}^{(2)}(t)}{R_{ext}.R_{m}^{2}C_{m}} = -c.I_{ephap}^{(2)}(t)
    \label{eq:ephaptic}
\end{equation}

\noindent
which corresponds to the ephaptic transmembrane current in the neuron (1) due to the ephaptic field of the neuron (2) [see figure \ref{fig:fig1}(a)]. Here, $R_{m}$ is the membrane resistance and $R_{ext}$ is the resistance of the extracellular milieu \cite{holt&koch1999,logothetis2007vivo,anastassiou2011,cunha2022ephaptic,cunha2023electrophysiological}. This model follows the electrical circuit approach (see Fig.\ref{fig:fig1}(b)).

\noindent
In this way, the QIF-E equation for ephaptic coupling is obtained by substituting  equation (\ref{eq:ephaptic}) into equation (\ref{eq:qif_adapt}). The case of the N-neuron system is calculated by applying the principle of superposition of electric potentials. The general QIF-E equations for the N-neuron system are therefore as follows:

\noindent
\begin{equation}
        \Dot{V}_{m}^{(i)}(t) = a . (V^{(i)}_{m}(t))^{2}+b . V_{m}^{(i)}(t) -\sum^{N}_{j \neq i} c_{(j)}'.(V_{m}^{(i)}(t)-V_{m}^{(j)}(t)) +I_{0}^{(i)}(t) 
        \label{eq:final}
\end{equation}
\noindent
where $V_{m}^{(i)}$ is the voltage in the i-th neuron membrane. In addition, the term $c_{j}'$ expresses the weight of the ephaptic coupling for the j-th neuron coupled to the i-th neuron. This term is different for each pair of neurons $(i,j)$ because the ephaptic coupling depends on the distance between the neurons.

\noindent
In addition, the term $I_{0}^{(i)}(t)$ corresponds to the synaptic input. In this work, the CUBA model was applied for synaptic modeling\cite{roth2009modeling,gerstner2014neuronal} which shows an exponential decay (see \cite{roth2009modeling,gerstner2014neuronal}):

\begin{equation}
       I_{0}^{(i)}(t) = \sum^{N}_{k\neq i} \omega^{(k)}.exp^{\frac{-(t-t_{0}^{(k)})}{T}}
\end{equation}
\noindent
where the synaptic intensity, $\omega^{(k)}$, is not equal to zero in the presence of a synaptic connection between the neurons $i$ and $k$. Furthermore, $t_{0}^{(k)}$ indicates the time at which the presynaptic (k) neuron had a spike. In addition, the parameters $a$ and $b$ are chosen in $a \in [25\pm1.25]$ and $b \in [30\pm1.5]$ to mimetic the biological difference between neurons.

In the next section, we discuss the complex network used in the simulation. We also introduce the QIF-E equation, which performs ephaptic coupling.

\subsection*{Complex Network model}

\noindent
The synaptic connection between the neurons follows a small-world topology \cite{watts1998collective}. The small-world network ensures that the information of the system has a short length and a high clustering coefficient \cite{watts1998collective,bassett2011understanding}. This topological arrangement is often used to describe the propagation of information in the brain, mainly because of its low energetic cost \cite{masuda2004global,liu2022analysis,bassett2011understanding,guardiola2000synchronization,sporns2022structure,bassett2006small}. These two properties are commonly associated with brain structure \cite{masuda2004global,liu2022analysis,bassett2011understanding,guardiola2000synchronization,sporns2022structure,bassett2006small}. In our study, most simulations were performed with $N =100$ neurons, with four first neighbors and a synaptic rewiring probability ($rp$) that varies from $0\%$ (regular network) to $100\%$ (aleatory network) (Fig.\ref{fig:fig1}(c) and (d), blue).

\noindent
To mimic the ephaptic network, the "all-to-all" topology was used, where the weighting is determined by the distance between neurons. Indeed, the ephaptic current depends on the electric field, which decreases quadratically with distance. Moreover, the electric field can be summed due to the superposition principle (Fig.\ref{fig:fig1}(d)).
In this work, the weights were estimated to the half of the network with an initial distance of $d = 50\mu m$ multiplied by the factor $\frac{1}{\vert i-k\vert}$, where $i$ and $k$ are the index of the neurons (Fig.\ref{fig:fig1}(b), red). To the other half of the network, the distances are estimated by the symmetry of the problem, using the network first half values. In this way, the ephaptic weight $c = \frac{10^{-2}}{\vert i - k \vert}$, where the dimension factor, $10^{-2}$ was previously estimated \cite{anastassiou2011,holt&koch1999,cunha2022ephaptic,cunha2023electrophysiological} using the physiological parameters described by Cunha et al.\cite{cunha2022ephaptic,cunha2023electrophysiological}.

The equation for a single neuron underlying the network is therefore as follows:
\begin{equation}
        \Dot{V}_{m}^{(i)}(t) = a . (V^{(i)}_{m}(t))^{2}+b . V_{m}^{(i)}(t) -\sum^{N}_{j \neq i} c_{(j)}'.(V_{m}^{(i)}(t)-V_{m}^{(j)}(t)) +\sum^{N}_{k\neq i} \omega^{(k)}.exp^{\frac{-(t-t_{0}^{(k)})}{T}}+I 
        \label{eq:qifecoupling}
\end{equation}

where the reset conditions are given by $V_{m}(t) \geq 90 mV \rightarrow V_{m}(t+1) = -5 mV$ (hyperpolarization condition).

\noindent
As for the numerical simulation, the time series was obtained from the average activity of all neurons in the network. To calculate each $V_{m}^{(i)}(t)$ in equation (\ref{eq:qifecoupling}), we use a time of $60$ s (excluding the first $10$ s as the transient phase), with a time step of $10^{-3}$ s. All simulations are performed in MATLAB using Euler integration with a step of $10^{-3}$ ($1ms$). In this way, the temporal series (Local Field Potential - LFP) is defined by the spatial average of all potential differences of the membranes in the networks, i.e:

\begin{equation}
    x_{l}(t) = LFP(t) = \frac{1}{N}\sum^{N}_{(i)} V_{m}^{(i)}
\end{equation}

The analysis presented in this paper refers to the time series $x_{l}(t)$, i.e., the activity average of the neuron networks (LFP). The QIF-E code are present in the Supporting Information. The parameter values listed in table \ref{tab:tab1} were used in our simulations.

\begin{table}[!h]
    \centering
    \caption{The model Parameters Values}
    \begin{tabular}{|c|c|c|}
    \hline
        Parameter & Value & Ref.\\
    \hline
    \hline
        $a$ & $25\pm1.25$ & \cite{cunha2022ephaptic,cunha2023electrophysiological}\\
         \hline
        $b$ & $30\pm1.5$ & \cite{cunha2022ephaptic,cunha2023electrophysiological}\\
         \hline
        $c'_{j}$ & $\frac{5.10^{-2}}{d. |i-k|}$ & \cite{cunha2022ephaptic,cunha2023electrophysiological,anastassiou2011,logothetis2007vivo}\\
        \hline
        $d$ & $50\mu m$& \cite{cunha2022ephaptic,cunha2023electrophysiological,anastassiou2011}\\
        \hline
        $T$ & 6ms & \cite{roth2009modeling,gerstner2014neuronal}\\
        \hline
        $I$ & 9.5 & \cite{cunha2022ephaptic,cunha2023electrophysiological}\\
    \hline
    \end{tabular}
    \label{tab:tab1}
\end{table}

\section*{Complexity and Multiscale Entropy}

\noindent
The entropy is a measure commonly associated with information to quantify complexity in systems \cite {shannon1948mathematical,zhang1991complexity,costa2002multiscale,richman2000physiological,humeau2015multiscale}. If we  consider $\textbf{X} := \{x_{1},x_{2},...,x_{M}\}$ a time series of length $M$, the sampling entropy ($S_{E}$) is a robust and widely used tool to quantify the regularity in $\textbf{X}$ \cite{richman2000physiological,costa2002multiscale,costa2008multiscale,humeau2015multiscale}. The samples $\bf{x_{m}}$ of $\bf{X}$ are defined for different $m$ lengths, i.e. $\bf{x_{m}} \subset \textbf{X}$, and $\textbf{x}_{m} = \{x_{l},x_{l+1},...,x_{m+l}\}$, with $1\leq m < M$. Note that $\textbf{x}_{m} \neq x_{m}$. $\bf{x_{m}}$ is the subset of $\textbf{X}$. Otherwise, $x_{m}$ is the $m$-th term of $\bf{x_{m}}$. The case $m=1$ therefore provides the original time series, $\bf{X}$, with $\textbf{x}_{m} = \{x_{m}\}$, $1\leq m \leq M$. The entropy of the sample is therefore defined by the following equation\cite{richman2000physiological,chenxi2016complexity,costa2002multiscale,costa2008multiscale,humeau2015multiscale}:

\begin{equation}
    S_{E}(m,r) = -ln\left( \frac{\sum^{M-m}_{l=1} n^{(m+1)}(r)}{\sum^{M-m}_{l=1} n^{(m)}(r)} \right)
    \label{eq:se}
\end{equation}
\noindent
where $n^{(m)}$ indicates the conditional probability that two samples $\bf{x_{m}}$ match in $m$ points. The agreement between the samples represents a tolerance $r$ that occurs. The tolerance match $r$ indicates the maximum Euclidean distance between two samples (of length $m$) in the same time series with the same $m$ values\cite{costa2002multiscale,costa2008multiscale,richman2000physiological,humeau2015multiscale}.

However, due to the limitations imposed by the usual random noise in experimental data, there is a problem in estimating signal complexity with strong random noise \cite{costa2002multiscale,costa2008multiscale,humeau2015multiscale}. Nevertheless, an information-poor signal can have a high entropy value due to the randomness of the data. To eliminate this problem, the entropy is estimated on different time scales to evaluate how the information is preserved along the time series. This approach is known as multiscale entropy (MSE) \cite{costa2002multiscale,costa2008multiscale,humeau2015multiscale}.

To compute the MSE, a coarse-graining procedure is used that generates new temporal series by applying the average to the subsets of size $\frac{M}{\tau}$ in the subsets of temporal series points \cite{costa2002multiscale,costa2008multiscale,humeau2015multiscale}. Note that the subsets in the coarse-grained method are not the same as the samples in the $S_{E}$ calculation. The consecutive number of points used in the procedure provides the scaling factor $\tau$, which generates the new replacement time series $y^{(\tau)}_{j}$ as the number of points.

\begin{equation}
    y^{(\tau)}_{j} =\frac{1}{\tau} \sum^{j\tau}_{k=(j-1)\tau+1} x_{k}
\end{equation}

\noindent
To obtain the MSE, a coarse-graining procedure is applied to different scale factors and a sample entropy is estimated for each new temporal series ($S_{E}(y^{(\tau)}_{j})$).

\noindent
In the complexity perspective, entropy is sampled across all time scales to estimate the information, variety, or randomness in the data \cite{costa2002multiscale, costa2008multiscale, zhang1991complexity,humeau2015multiscale}. The system complexity, \textit{K}, is thus defined by calculating the entropy integral along the time scales $\tau$: \cite{zhang1991complexity}.

\begin{equation}
    K = \int_{1}^{\tau_{max}} S_{E}(y^{(\tau)}) d\tau
\end{equation}

where $\tau_{max}$ is the maximum $\tau$ value that matches the data size\cite{humeau2015multiscale}. In the present work, we use a temporal series with length $M = 50000$, $\tau \in [2,100]$, $m=2$ and $r = 0.15*SD$, for \textit{SD} the temporal series standard deviation.
% Results and Discussion can be combined.
\section*{Results}

%-CORRECAO GRAMATICAL AQUI-

\begin{figure}[H]
    \centering
    \includegraphics[width=1\textwidth]{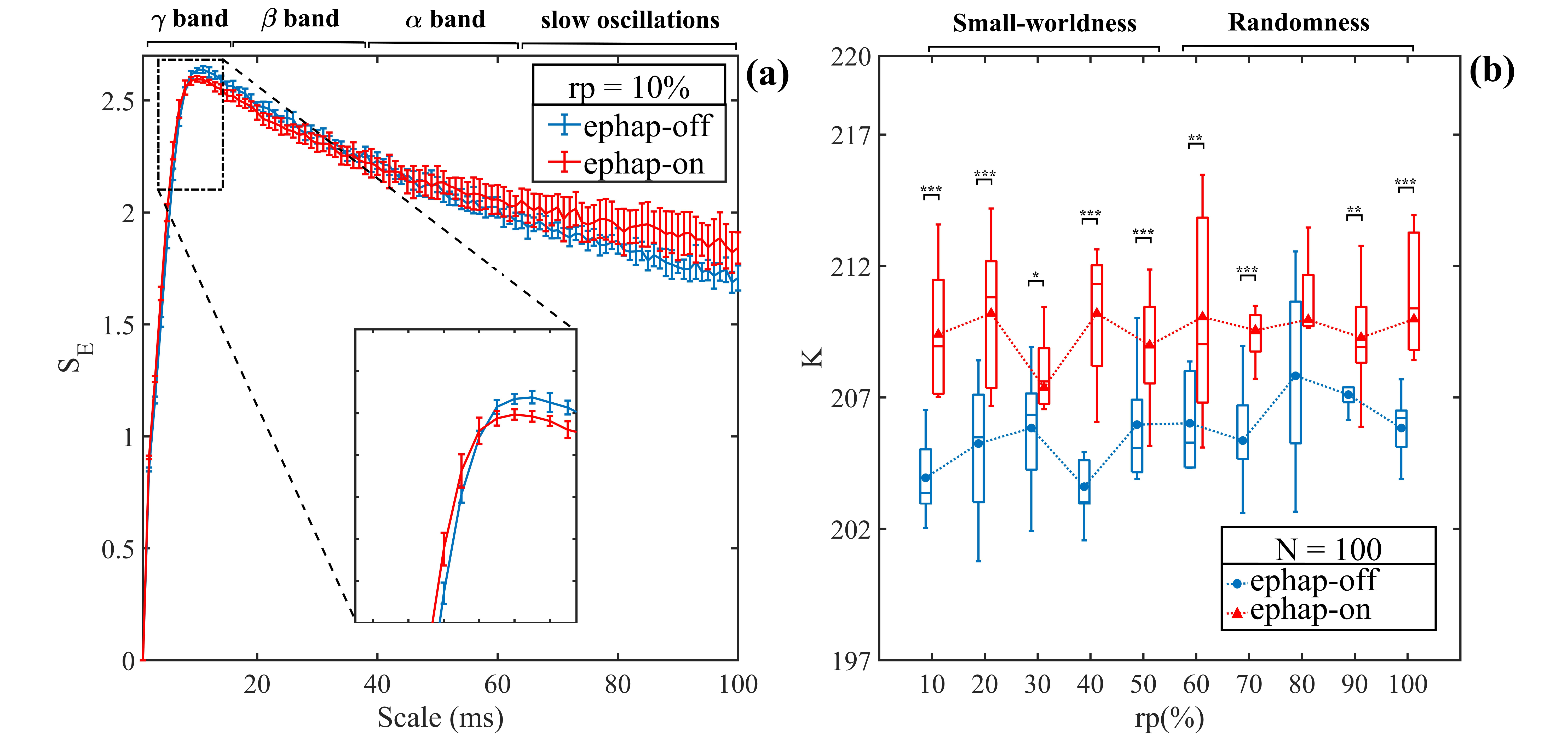}
    \caption{\textbf{MSE method and complexity measure for the two models: ephaptic-off and ephaptic-on networks} (a) Multiscale entropy for the ephaptic-off network LFP ($rp= 0.1$ and $\omega^{(k) } = 5$, in blue) and the  ephaptic-on network LFP ($rp= 0.1$ and $\omega^{(k)} = 5$, in red). Both models were simulated for N=100. (b) Complexity ($K$) for different rewiring probabilities (\textit{rp}) in the small-world ephaptic-off network. All these simulations were performed with N = 100. The colors are the same as in panel (a). Wilcoxon rank sum test was applied ($* \rightarrow p < 0.05$, $** \rightarrow p < 0.01$, $*** \rightarrow p < 0.001$).}
    \label{fig:fig2}
\end{figure}

In the present section, we show the results of the simulations for both models used: ephaptic-off and ephaptic-on. Using the MSE tool, we compute the complexity in networks with small worlds. Figure \ref{fig:fig2}(a) shows the MSE for the ephaptic-off network (small world; rewiring probability (rp) = $10\%$) and ephaptic-on (synaptic: small world (rp $= 10\%$)). We note that at short scales (8 ms), the  ephaptic-on network (Fig.\ref{fig:fig2}(a), in red) is more entropic than the ephaptic-off network (blue). On the other hand, for the intermediate range (from 8 ms to 40 ms), the ephaptic-off network exhibits greater MSE than the ephaptic-on system. At long range scales (from 40 ms to 100 ms), the  ephaptic-on network became more entropic compared to the pure ephaptic-off network. In general, the complexity estimated by the MSE integral (area below curves) is larger in ephaptic-on networks (in red) than ephaptic-off (in blue) observed in Fig. \ref{fig:fig2}(b), for different rewiring probabilities (\textit{rp}) in the small-world topology. To confirm these results, the Wilcoxon rank sum test ($* \rightarrow p < 0.05$, $** \rightarrow p < 0.01$, $*** \rightarrow p < 0.001$) was performed comparing networks with the same \textit{rp} values. However, it can be observed that $rp = 80\%$ shows no significant difference between the two cases.

\begin{figure}[H]
    \centering
    \includegraphics[width=.92\textwidth]{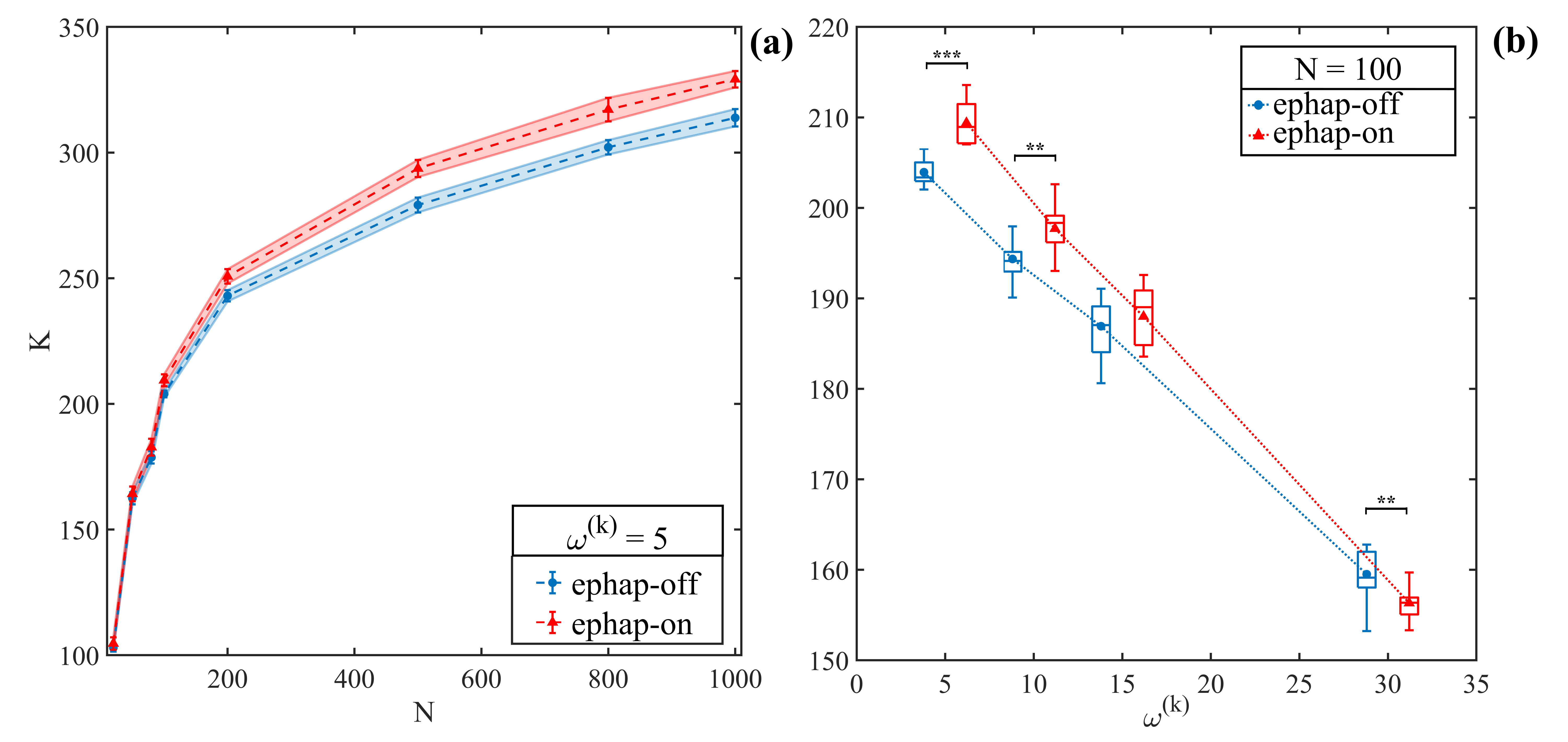}
    \caption{\textbf{Complexity for ephaptic-off and ephaptic-on Networks.}(a) Complexity for different neuron quantities, with $rp= 0.1$ and $\omega^{(k)} = 5$ (ephaptic-off network, in blue; ephaptic-on network, in red). Both models were simulated for N=100. (b) Complexity for different synaptic intensity. The increase of $\omega^{(k)}$ promotes an decrease in complexity (K). The behavior is common both systems (ephaptic-off and -on). The complexity for weaky synapses is amplified by addition of ephaptic process ($\omega^{(k)} = 5$ and $\omega^{(k)} = 10$). Otherwise, to strong synapses, the ephaptic-on system have an decrease comparing with analogous ephaptic-off ($\omega^{(k)} = 30$). Wilcoxon rank sum test was applied ($* \rightarrow p < 0.05$, $** \rightarrow p < 0.01$, $*** \rightarrow p < 0.001$).}
    \label{fig:fig3}
\end{figure}

%\noindent
Subsequently, topological simulations were performed for several numbers of neurons, denoted as N, to explore the behavior of the model for increasing N.  Figure \ref{fig:fig3}(a) illustrates the complexity of the small-world ephaptic-off network ($rp = 10\%$, shown in blue) and the ephaptic-on network ($rp = 10\%$, shown in red). It is evident that the ephaptic-on network has greater complexity than the ephaptic-off network. Furthermore, the significant discrepancy between the average curves occurs from N=100 and increases with the inclusion of neurons in the network. Both average complexity curves show a simultaneous decrease in the slope of the curve for large N.

%\noindent
To explore the behavior of complexity in response to variations in synaptic weights ($\omega^{(k)}$), in Figure\ref{fig:fig3}(b) we investigate the effect of synaptic strength. Our results reveal that as the increase in synaptic strength occurs, it results in a reduction in the complexity of the network, regardless of whether the ephapticity is turned on or not. For lower synaptic intensities ($\omega^{(k)} <$  20) the complexity of the ephaptic-on network (red) is higher than that of the ephaptic-off network (blue). In contrast, at higher synaptic intensities ($\omega^{(k)}>$ 20) the complexity of the   ephaptic-on network (red) exhibits a lower value when compared to the complexity value of the ephaptic-off network (blue). The Wilcoxon rank sum test was applied to complexity data, K is larger for the ephaptic-on network from most of the cases. Only  for $\omega^{(k)} = 15$ there is no difference in network complexity ($* \rightarrow p < 0.05$, $ ** \rightarrow p < 0.01$, $*** \rightarrow p < 0.001$). It is notable that the complexity is greater for ephaptic-on structures at $\omega^{(k)} = 10$. However, for high synaptic strength, $\omega^{(k)} = 30$, the ephaptic-off network presents higher complexity than the combined network (ephaptic-on).

\begin{figure}[H]
    \centering
    \includegraphics[width=1.02\textwidth]{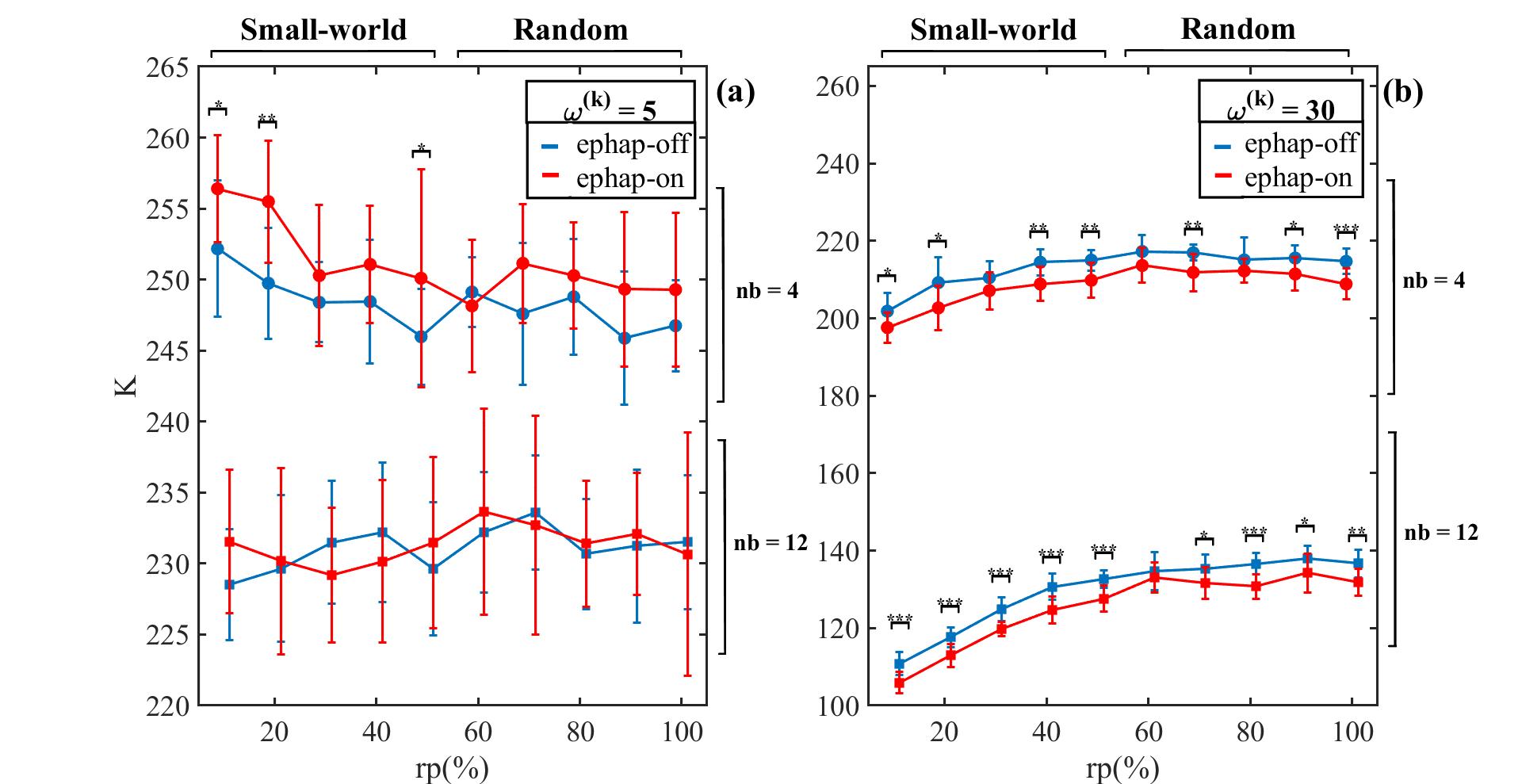}
    \caption{\textbf{Complexity for N=200, with different neighborhoods.} (a) The mean value for complexity measure for network with $\omega^{(k)}=5$. Note that the number of neighbors impact the complexity. However, for $nb = 4$, the difference between ephaptic-off and ephaptic-on networks is higher compared to $nb = 12$. Furthermore, in the case of $rp < 50\%$  the complexity is smaller compared to $rp > 50\%$. (b) The same comparison that in (a), for $\omega^{(k)}=30$. The values of complexity for $\omega^{(k)} = 30$ is inferior in comparison with the $\omega^{(k)} = 5$ (fig.(a)). In addition, the complexity is higher between neighborhoods than fig.(a). Moreover, the ephaptic-off networks exhibit larger complexity that ephaptic-on networks, see fig. \ref{fig:fig3}(b).  Wilcoxon rank sum test was applied ($* \rightarrow p < 0.05$, $** \rightarrow p < 0.01$, $*** \rightarrow p < 0.001$).}
    \label{fig:fig4}
\end{figure}

%\noindent
Finally, figure \ref{fig:fig4} shows the results of complexity for different neighbor number (nb) and synaptic intensity, in networks with $N=200$. Figure \ref{fig:fig4}(a) shows complexity for $\omega^{(k)} = 5$. In the weakly ephaptic-off regime, the neighbor number inversely impacts the complexity. In other words,  increasing the  number of neighbor decreases the complexity of the dynamics in the network. Furthermore, the ephaptic coupling increases its complexity significantly for rp $\leq $30\ (small-world) and only for $nb = 4 $ Thus, for $nb=12$, the results do not change significantly between synaptic and ephaptic-on networks for $nb=12$. For $nb=12$, the results do not change significantly between ephaptic-off and ephaptic-on connections. In Figure 4b, the complexity of strong synaptic connections is shown.  As we observed, for the strongest synaptic connection, the turn-on ephapticity (red) leads to a lower system complexity than without ephapticity  (blue), regardless of the number of neighbors (\textit{nb}) and the probability of reconnection (\textit{rp}). Also, the significant difference between the   networks appears for almost all \textit{rp} and \textit{nb} (take into account the statistical fluctuations). Note that increasing neighbor number decreases  complexity (See also Supporting Information). However, $nb = 12$ exhibits less complexity than $nb = 4$. The Wilcoxon rank sum test was applied to complexity data ($* \rightarrow p < 0.05$, $ ** \rightarrow p < 0.01$, $*** \rightarrow p < 0.001$)

\section*{Discussion}

In this study, using the QIF-E hybrid model \cite{cunha2022ephaptic}, we analyzed the effect of ephaptic coupling on neural complexity in synaptic networks (ephaptic-off). We used the average time series (LFP) of the network in question to calculate the multiscale entropy on a scale that varies from 2 to 100 ms \cite{costa2002multiscale, costa2008multiscale}. In order to model the ephaptic-off network, we chose a small-world topology, recognized as the most suitable network configuration to represent complex neuronal functioning on multiple time scales.\cite{watts1998collective, masuda2004global, liu2022analysis, bassett2011understanding, guardiola2000synchronization, sporns2022structure,bassett2006small}. The ephaptic-on model was incorporated into the network [see Fig.\ref{fig:fig1}]. The results revealed  a significant enhancement in neuronal complexity by incorporating ephaptic coupling as an underlying mode of communication in the synaptic nervous system \cite{ks1940, arvanitaky1942, holt&koch1999, anastassiou2011, cunha2022ephaptic}. The discoveries indicate that ephaptic interactions could be crucial in fine-tuning the complexities of the neural network, providing valuable perspectives on intricate dynamics across multiple levels. Our findings are in harmony with recent empirical research \cite{ryan2015engram, billings2018disentangling, wang2018neurophysiological, hunt2023fields,van2023now}.

The results depicted in Figure \ref{fig:fig2}(a) reveal that within intervals shorter than 10 ms (short range) and longer than 20 ms (long range), the average entropy of the ephaptic-off (blue) network experiences a significant rise upon the introduction of ephaptic coupling (red). Courtiol et al. (2016) \cite{courtiol2016multiscale} demonstrated that this form of neuronal communication enhances complexity across both high and low frequencies, which is consistent with our findings. Conversely, it is observed that the ephaptic-off (blue) network exhibits, on average, a slightly higher entropy value only in the medium time scales, within a small range from 10 ms to 15 ms, compared to the ephaptic-on network. Consequently, oscillations in intermediate bands of the spectrum, between the high $\beta$ ($13-30$ Hz) and low $\gamma$ ($45- 100$ Hz) bands, are related to a possible reduction in complexity when the ephaptic process is turned on \cite{courtiol2016multiscale}. The scales in the MSE in Figure \ref{fig:fig2}(a) are associated with different neural oscillations, according to Coutiol et al. \cite{courtiol2016multiscale}.  These results suggest that various modes of communication can take on specific roles depending on the specific neuronal task at hand, thus exerting an influence on the oscillation patterns observed in the neural signal \cite{berridge2014calcium, palva2011functional, wang2010neurophysiological, fell2011role}. 

When evaluating the complexity of the neural network for N = 100, Figure \ref{fig:fig2}(b) shows that topologies associated with different synaptic reconnection rates ($rp\%$) have no influence on the average complexity of the networks. This observation persists as an intrinsic statistical fluctuation in the randomness of the ephaptic-off network reconnection process. In conclusion, the results in Figure \ref{fig:fig2}(b) suggest that, on average, network complexity is greater when neuronal communication involves ephaptic coupling, regardless of the probability of reconnection in the ephaptic-off network, for the case $N=100$.

The complexity of both networks regimes increases with the addition of new neurons, as evidenced in Figure \ref{fig:fig3}(a). An important observation is the marked difference between the two regimes (curves), especially for $N>100$, showing always a greater degree of complexity when ephapticity is turned on. Indeed, this discrepancy amplifies with the escalation in neuron number, suggesting a potential contribution of ephaptic coupling to large-scale brain complexity ($N > 200$). The idea that small-world network models become more complex as  the number of neurons increases is driven by various interrelated factors. Even in an ephaptic-off (synaptic) regime, studies such as Song et al. (2005) \cite{song2005highly} and Bullmore and Sporns (2012) \cite{bullmore2012economy} highlight the importance of neuronal interconnections in forming complex networks. As the number of neurons increases, there is an expansion in synaptic connection opportunities (as well as ephaptic too), leading to a denser and more complex network of neuronal communication. This enables greater integration of information, as discussed by Tononi et al. (1994) \cite{tononi1994measure} and Sporns et al. (2004) \cite{sporns2004organization}, contributing to the representation and processing of more complex information in the brain. 

The increase in the number of neurons can lead to the emergence of more intricate patterns of neuronal activity, as described by Buzsáki and Draguhn (2004) \cite{buzsaki2004neuronal}, adding complexity (plasticity) to the system. In short, the complexity of the brain may be a consequence of the increase in the number of neurons, reflecting the intricate and highly organized nature of the nervous system. This complexity is driven by expanded neuronal interconnection, enhanced information integration, the emergence of complex activity patterns, functional diversity, and neuronal plasticity, all of which combine to create a complex, dynamic neural system. The small-world neuronal model may be under the same process.

On the contrary to what was observed in Figure\ref{fig:fig3}(a), Figure \ref{fig:fig3}(b) exhibits a decrease in complexity as the intensity of the synaptic interaction increases. We defined four synaptic strengths to simulate the complexity, ranging from the weakest synaptic strength $\omega^{(k)} = 5$ to a robust synaptic strength of $\omega^{(k)} = 30$, for $N=100$. The results reveal an inversely proportional correlation between synaptic intensity ($\omega^{(k)}$) and complexity in both analyzed network regimes.Furthermore, in the ephaptic-on regime, the complexity significantly exceeds the complexity of the ephaptic-off regime, particularly for weak synaptic connections. This observation suggests a potential strategy to modulate the balance between synaptic and ephaptic communications and perhaps enhance the efficiency of a stable-state (healthy) nervous system. This conclusion resonates with the one in Figure \ref{fig:fig2}(a) regarding the various oscillatory modes prevalent in the network signal and their associated maximum complexity timescales. Therefore, efficacy and ephaptic coupling can coexist harmoniously with synaptic communication  to achieve an optimized state of complexity in the brain. Our results also highlight an even more accentuated reduction in complexity for the ephaptic-on regime compared to the ephaptic-off regime, as the intensity of synapses is strong ($\omega^{(k)} = 30$).

To provide further context for our findings regarding the outcomes depicted in Figure \ref{fig:fig3}(b), an intriguing study by Iara B. et al. \cite{Iara2024} employed the concept of recurrence entropy via microstates \cite{corso2018quantifying} as a metric for complexity. They demonstrated a decrease in entropy during the eyes-closed condition, indicating that intrinsic neural activity within the thalamocortical circuit predominates, characterized by the alpha rhythm oscillating between 8 and 12 Hz. Conversely, in the eyes-open condition, neural responses in the occipital region, responsible for visual processing, lead to desynchronized neuronal activity due to light detection on the retina. In another study, Asl M. et al. \cite{madadi2022spike} elucidated spike timing-dependent plasticity as a fundamental neural mechanism, altering synaptic strengths based on the temporal alignment of pre- and postsynaptic spikes \cite{gerstner1996neuronal}. In Parkinson's disease, for instance, dopamine depletion can induce dysfunction reliant on synaptic connectivity, fostering pathological states characterized by tightly synchronized activity in closely connected neurons, indicative of diminished plasticity and complexity \cite{hammond2007pathological}. Additionally, computational investigations have delved into the relation between synchronization and desynchronization dynamics of neurons and their firing patterns, contingent upon factors such as network topology, neuronal population size, and coupling strength \cite{belykh2005synchronization}, thereby reinforcing our analyses. It's noteworthy that these studies exclusively focus on regimes involving synaptic interactions, without considering ephaptic communication.

%-9
In order to elucidate the effects of neuronal complexity and relate them to the causality of ephapticity, we show that our findings corroborate existing empirical studies that highlight the importance of ephapticity in the organization and complexity of the brain. It is now well established that neural activity, manifested as waves or spikes, traditionally propagates through mechanisms such as synaptic transmission, gap junctions, or diffusion. However, the paper proposed by Chen Qiu (2015) \cite{qiu2015can} elucidated an alternative explanation for experimental data suggesting that neural signals may propagate via an electric field mechanism, known as ephaptic effects, to mediate the propagation of self-regenerating neural waves. This novel mechanism, involving cell-by-volume conduction, could potentially play a role in various types of propagating neural signals, including slow-wave sleep, sharp hippocampal waves, theta waves, or seizures. Another study demonstrated that ephaptic interaction alone can shape circuit function, inducing lateral inhibition of neurons \cite{zhang2019asymmetric}. This interaction influences spike timing, facilitating the development of intricate neural codes in higher processing centers. Moreover, in this study, the authors showed that neurons with large spikes can exert greater ephaptic influences on their neighbors and potentially display ephaptic asymmetry, potentially stemming from an unequal number of reciprocal synapses, may leading to a preference for the ephaptic pathway.

Hence, while specific cerebral mechanisms may differ, it is plausible that ephaptic dynamics regulate and modulate certain aspects of brain information, especially over synaptic processes during the advanced stages of neurodegenerative diseases \cite{cunha2023electrophysiological}. This abnormal scenario intensify the reduction in complexity, evident in Fig.\ref{fig:fig3}(b), may attributable to the inflexibility (rigidity) induced by strong synaptic connections. Loss of physiological functionality, a hallmark of the degenerative process, could be the cause of the ephaptic asymmetry and synaptic rigidity, affecting a specific subset of neurons in a specific region of the brain. The intricate interplay between synaptic and ephaptic processes, more specifically, in neurodegenerative contexts, underscores the need for a nuanced understanding of how disruptions in these mechanisms contribute to the observed alterations in brain dynamics. This insight is vital for unraveling the complexity associated with neurodegeneration, and may pave the way for targeted interventions aimed at mitigating the impact of these diseases on neural networks. As we delve deeper into the intricate dynamics of synaptic and ephaptic interactions, potential avenues for therapeutic interventions are opened, offering hope for improved management and treatment strategies for individuals affected by neurodegenerative conditions.

Figure \ref{fig:fig4} highlights the small-world characteristics present in neuronal data \cite{masuda2004global, liu2022analysis, bassett2011understanding, guardiola2000synchronization, sporns2022structure, bassett2006small}. In Figure \ref{fig:fig4}(a), for $nb = 4$, it is notable that the complexity undergoes a transition at $rp \approx 50\%$. According to the Watts-Strogatz approach, networks with $rp$ less than $50\%$ predominantly exhibit small-world characteristics \cite{watts1998collective}. However, networks with $rp$ greater than $50\%$ demonstrate characteristics of random networks. Thus, the topological structure may influence the outcomes of our study, suggesting that communication balance depends on neuronal topological properties, which is supported by other studies \cite{masuda2004global, liu2022analysis, bassett2011understanding, guardiola2000synchronization, sporns2022structure, bassett2006small}.

In Figure \ref{fig:fig4}(b), it is observed that strong synaptic intensity decreases the network complexity, potentially fostering a synchronized environment among neurons in the network. Furthermore, the figure indicates that regardless of the reconnection probability, increased the synaptic intensity the ephapticity further reduces the system's complexity. These results are consistent with those in Figure \ref{fig:fig3}(b), suggesting that ephaptic communication exhibits causal dynamic behavior and plays a modulatory role within the synapse neural network.

Lastly, the relationship between complexity and the first synaptic neighbors shown in Figures \ref{fig:fig4} (a) and (b) provides insight into the interplay between synaptic and ephaptic communication. The decrease in complexity for $nb= 12$ is observed in both sets of synaptic strength results. Thus, we demonstrate that the findings in Figure \ref{fig:fig4} underscore the balance between ephapticity and neural synapses in brain communication structure, proposing a modulatory role of ephapticity in the central nervous system. These outcomes are in agreement concerning the effect of ephaticity in a strong synapse regime, from the study by Kyung-Seok Han and co-workes (2018) \cite{han2018ephaptic} that showed how ephatic coupling promotes synchronous firing of cerebellar Purkinje cells. 

These studies complement ongoing research and may shed light on the neurophysiological processes underlying brain complexity and their relationship with functional connectivity  \cite{wang2018neurophysiological}. The heightened complexity indicates enhanced organization and more frequent transitions between diverse integrative states within brain networks, as evidenced by Billings (2018) \cite{billings2018disentangling}. This prompts the hypothesis that ephaptic coupling in vivo might play a pivotal role in memory formation and consolidation, as demonstrated by Pinotsis (2023) \cite{pinotsis2023vivo}.

\section*{Conclusion}

Our research using the QIF E model revealed an interesting finding; the complexity curve generated by ephapticity activation seems to align with that produced solely through synaptic communication. This alignment suggests that synaptic processes may play a guiding role, in shaping the complexity of networks. This correlation was consistent across analyses including factors like network size, strength of synaptic connections number of connected neighbors and probability of synaptic reconnection.

Although ephaptic communication is significantly weaker (about 1000 times less intense) than synaptic communication, its activation increases system complexity by 7–13\%, depending on the state/regime observed. Therefore, our findings indicate that ephapticity is not merely insignificant background noise caused by synaptic activities in the network. Ephapticity exerts a significant influence on neuronal communication and the brain's energetic cost. This implies that it requires the same amount of energy as a pure synaptic network to improve the organization of neural connections and to increase the complexity of the same number of synaptic connections.

Our research suggests that ephapticity is a contributing factor  to the enhancement of complexity in certain neural network scenarios. Implementing ephaptic communication into models can produce a better understanding of how brain structure organizes communication. This communication may be associated with the interaction between nerve cells, their various states, and the effects given by cognitive and memory processes. Moreover, our findings provide new insights into the implications of ephaptic communication in neurodegenerative disorders that affect plasticity, generating an imbalance in the dynamics of central nervous system communication. It is possible that synaptic communication alone is not sufficient to convey the full picture of brain complexity, underscoring the potentially significant role of ephaptic communication.

The QIF-E hybrid neuronal network model offers a good perspective on analyzing brain complexity by integrating both synaptic and ephaptic communication mechanisms. One of its strengths lies in capturing the intricate interplay between these two forms of neural communication, which are increasingly recognized as crucial in understanding brain function. By incorporating ephaptic effects alongside traditional synaptic transmission, such models can better simulate real-world neural dynamics, potentially revealing emergent properties and behaviors not observable in purely synaptic models. This hybrid approach enables researchers to explore how ephaptic interactions modulate neural activity and contribute to overall brain complexity.

However, this approach also has its limitations. One major challenge is accurately modeling the biophysical properties of ephaptic coupling, as these interactions are less well understood compared to synaptic transmission. Additionally, incorporating ephaptic effects into network simulations increases computational costs, potentially requiring significant computational resources and time. As a result, researchers may need to make trade-offs between model complexity and computational feasibility, limiting the size of the network and the duration of simulations. Furthermore, determining the appropriate size and scale of the network is crucial but challenging. Scaling up the network to represent the entire brain (may N$\gg$1000)  introduces computational challenges and may require simplifications or approximations that limit model accuracy. Conversely, modeling small-scale networks (may N$<$100) may overlook emergent properties and complex interactions that are only manifested at larger scales. Balancing these considerations while maintaining biological plausibility and computational efficiency poses a significant obstacle to hybrid neuronal network modeling.

In conclusion, QIF-E hybrid neuronal network models offer valuable insights into brain complexity by integrating synaptic and ephaptic communication mechanisms. While these models provide a powerful tool for studying neural dynamics, they also face challenges related to accurately representing ephaptic interactions, computational complexity, and determining the appropriate scale of network simulations. Addressing these limitations will be crucial for advancing our understanding of brain function using hybrid neuronal network models.

\section*{Supporting information}

\subsection*{Complementary Results}

\begin{figure}[H]
    \centering
    \includegraphics[width=1\textwidth]{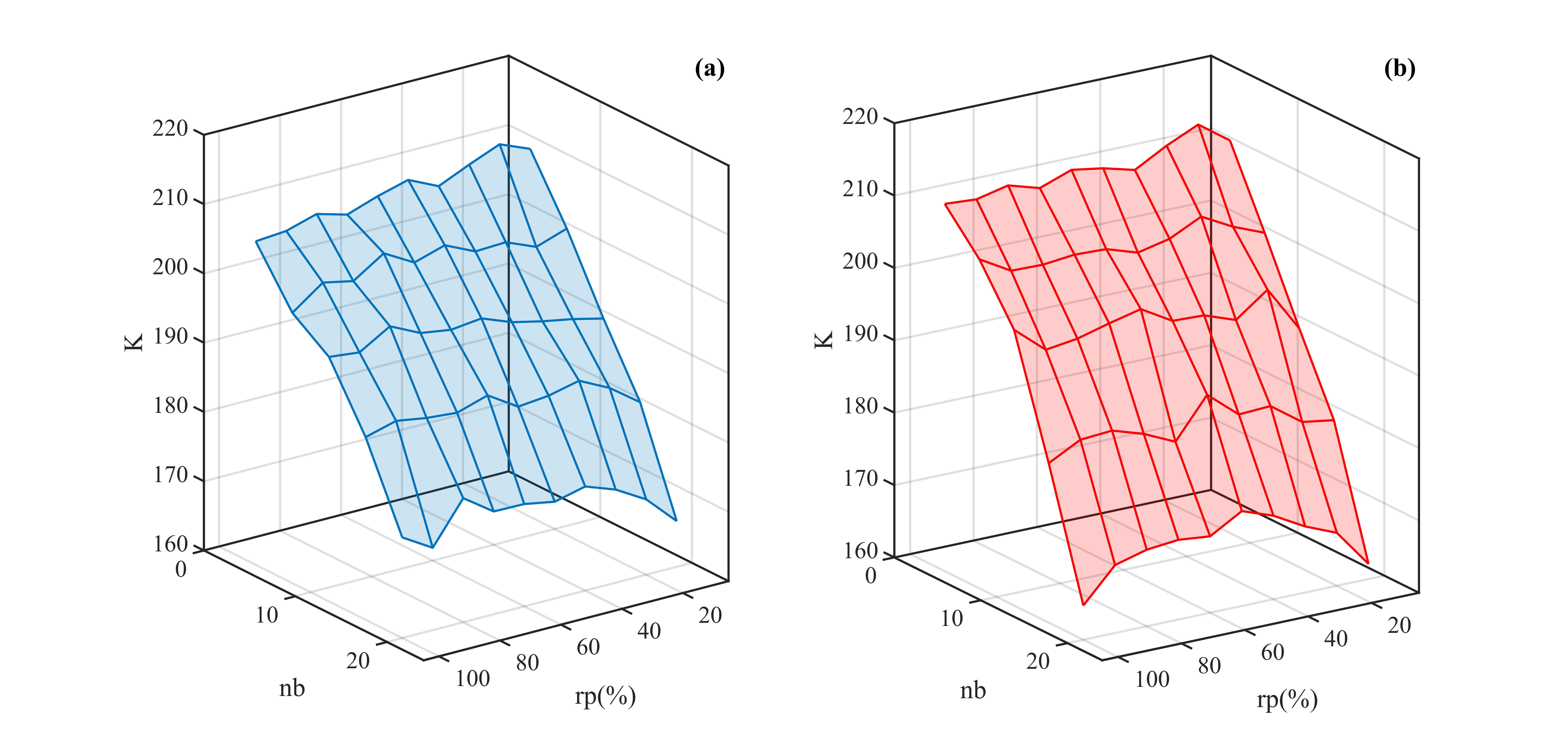}
    \caption{\textbf{Complexity for all neighborhoods and rp values performed, with $\omega^{(k)} = 5$ and $N=100$.} (a) Complexity (K) for different neighbors number (nb) and different rewiring probabilities (rp) values, to networks ephaptic-off small-world. Note that, to values of $rp \approx 10\%$, and $nb = 4$, the small-world features promote highest complexity. Otherwise, to $nb = 20$ the highest values of complexity are presented in random networks structure. (b) Complexity (K) for different neighbors number and different rp values, to networks ephaptic-on. Observe that the complexity values is highest than figure S1(a) in low $nb$ values. However, the increase of $nb$ promotes an acentuated decrease in comparison with (a). Therefore, the small-world prevalence occurs in minors $nb$.}
    \label{fig:S1_Fig}
\end{figure}

\begin{figure}[H]
    \centering
    \includegraphics[width=1\textwidth]{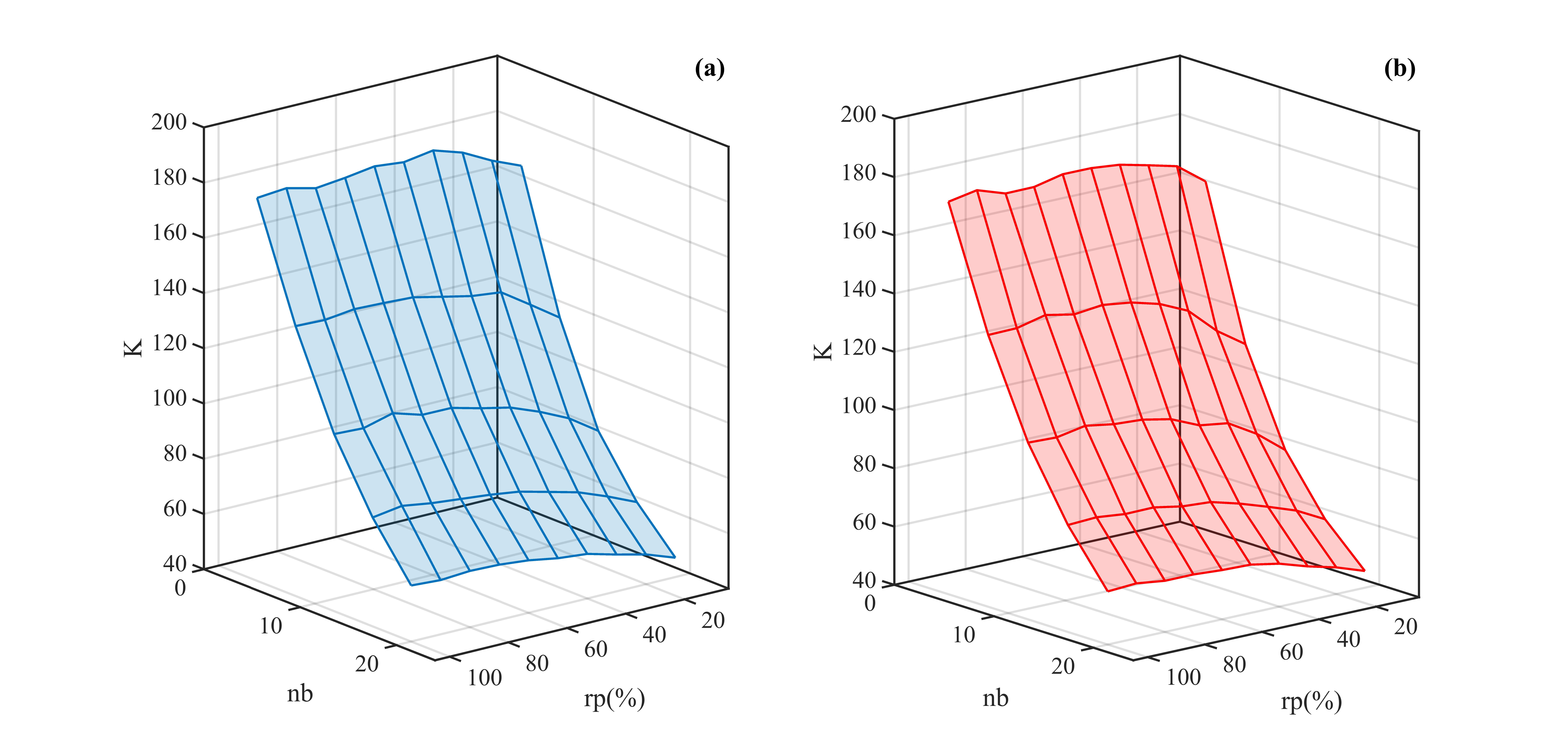}
    \caption{\textbf{Complexity for all neighborhoods and rp values performed, with $\omega^{(k)} = 30$ and $N=100$.} (a) Complexity (K) for different neighbors number (nb) and different rewiring probabilities (rp) values, to networks ephaptic-off small-world. The complexity to strong synapses is lower in comparison with figure \ref{fig:S1_Fig}(a), as shows by figures \ref{fig:fig3}(b) and \ref{fig:fig4}. (b) Complexity (K) for for different neighbors number (nb) and different rewiring probabilities (rp) values, to networks ephaptic-on. The values of complexity to combined networks is lower in comparison with (a). This results are in line with the figures \ref{fig:fig3}(b) and \ref{fig:fig4}.}
    \label{fig:S2_Fig}
\end{figure}

\begin{figure}[H]
    \centering
    \includegraphics[width=1\textwidth]{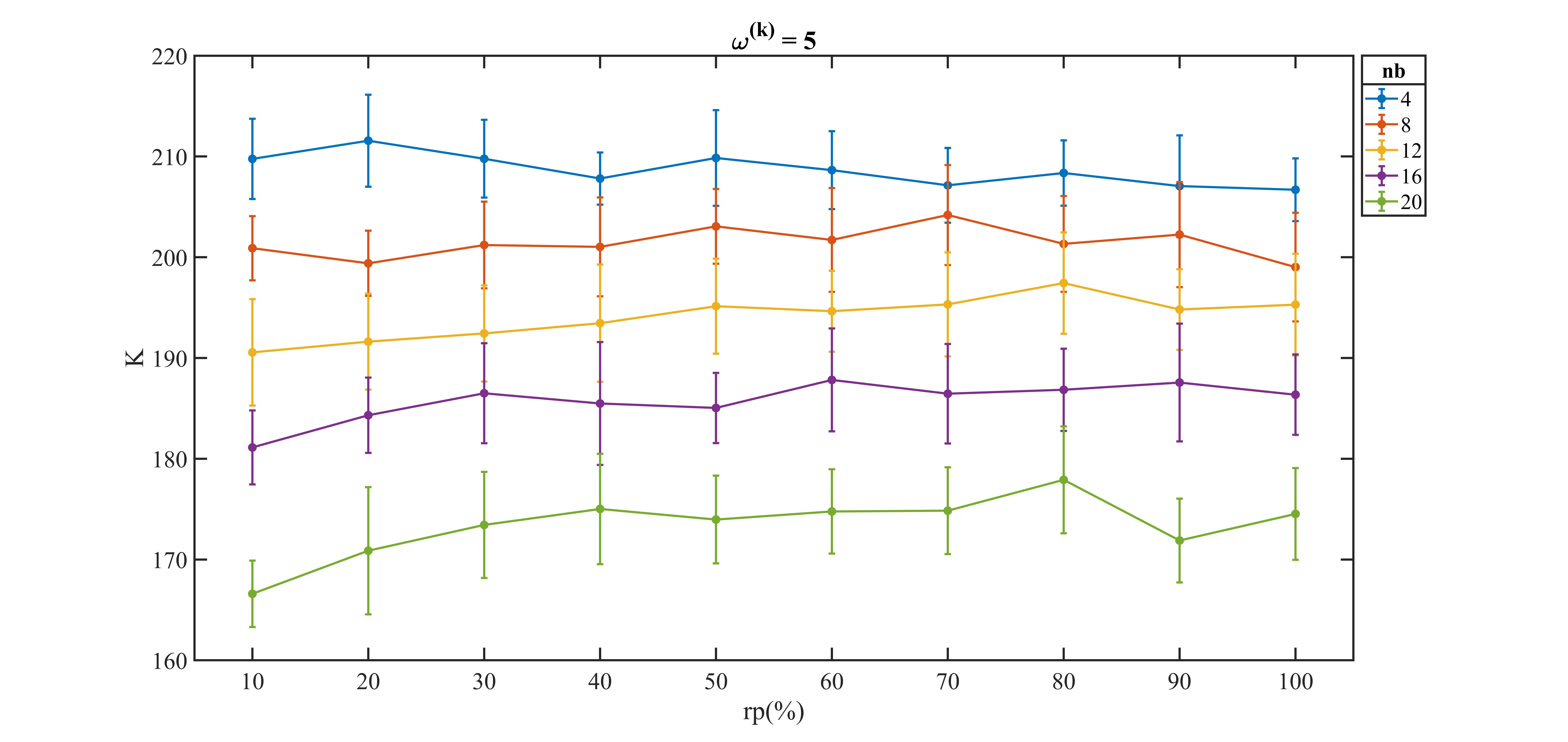}
    \caption{\textbf{Plane Projection of Figure \ref{fig:S1_Fig}(a).}}
    \label{fig:S3_Fig}
\end{figure}

\begin{figure}[H]
    \centering
    \includegraphics[width=1\textwidth]{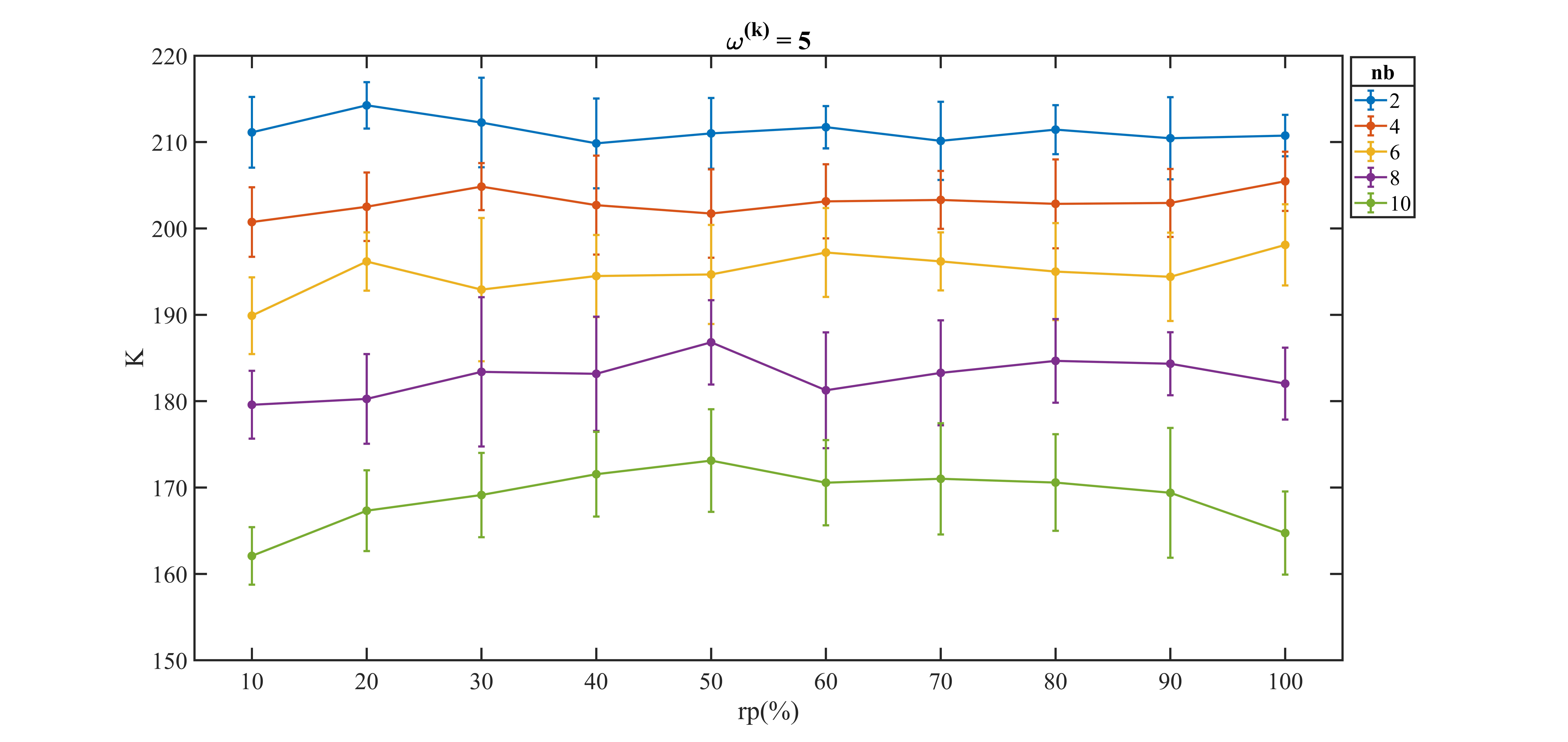}
    \caption{\textbf{Plane Projection of Figure \ref{fig:S1_Fig}(b).}}
    \label{fig:S4_Fig}
\end{figure}

\begin{figure}[H]
    \centering
    \includegraphics[width=1\textwidth]{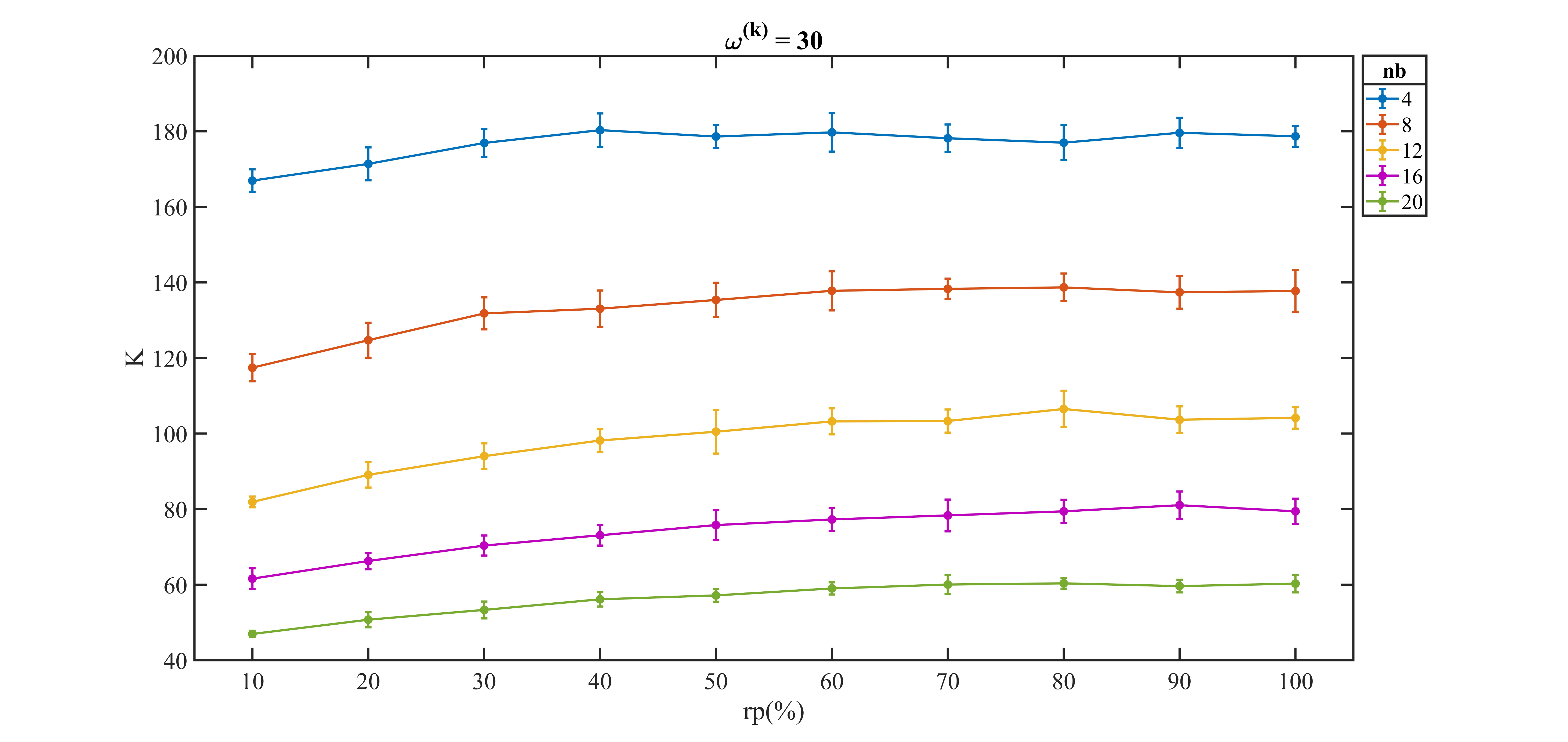}
    \caption{\textbf{Plane Projection of Figure \ref{fig:S2_Fig}(a).}}
    \label{fig:S5_Fig}
\end{figure}

\begin{figure}[H]
    \centering
    \includegraphics[width=1\textwidth]{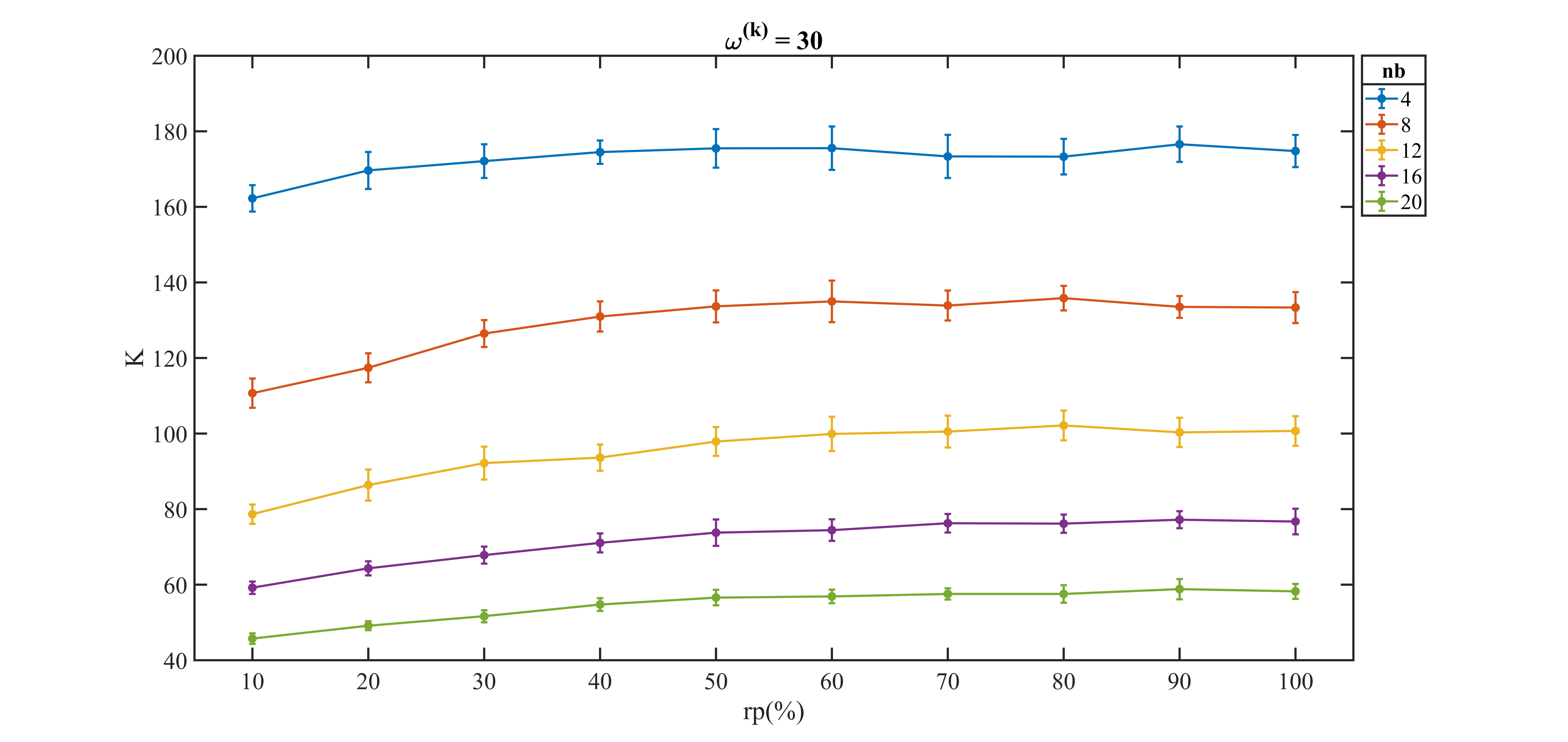}
    \caption{\textbf{Plane Projection of Figure \ref{fig:S2_Fig}(b).}}
    \label{fig:S6_Fig}
\end{figure}

\subsection{QIF-E Model Code}

\definecolor{codegreen}{rgb}{0,0.6,0}
\definecolor{codegray}{rgb}{0.5,0.5,0.5}
\definecolor{codepurple}{rgb}{0.58,0,0.82}
\definecolor{backcolour}{rgb}{0.95,0.95,0.92}

\lstdefinestyle{mystyle}{
    backgroundcolor=\color{backcolour},   
    commentstyle=\color{codegreen},
    keywordstyle=\color{magenta},
    numberstyle=\tiny\color{codegray},
    stringstyle=\color{codepurple},
    basicstyle=\ttfamily\footnotesize,
    breakatwhitespace=false,         
    breaklines=true,                 
    captionpos=b,                    
    keepspaces=true,                 
    numbers=left,                    
    numbersep=5pt,                  
    showspaces=false,                
    showstringspaces=false,
    showtabs=false,                  
    tabsize=1
}

\lstset{style=mystyle}
\begin{lstlisting}[language=matlab]
% Programme designed to simulate ephatic coupling, using the
% ephatic quadratic integrate-and-fire model.
% Developed on 12/2023 by Gabriel Moreno Cunha.
% Neurodynamic Simulation and Modelling Laboratory-UFRN, Natal-RN-BR
%% ------------
% INPUT PARAMETERS OF THE FUNCTION (IN THE ORDER OF THE INPUT):
%A. n -> Number of Neurons (Greater than 2)
%B. viz -> Half Number of neighbours
%C. simulate -> Number of simulations for a set of fixed parameters
%(integer value)
%D. t_final -> Total time in seconds
%E. prob -> Probability of reconnection in the synaptic small-world (less than 1)
%F. weight -> synapse intensity
%G. pesoefa -> exponent of ephatic communication (USE 0 IF YOU DON'T WANT TO APPLY)

%% ------------
% OUTPUT PARAMETERS OF THE FUNCTION (IN THE ORDER OF THE OUTPUT WITH DIRECT SAVE):
%A. v -> LFP of the network
% Enjoy of it!
clear all
%Simulation parameters
n = 50; % neuron number
viz=2; % Half Neighbor number
simul=[1]; % Number of simulations
t_final=10; % Each simulation Total Time (s)
prob=1; % Rewiring probability
peso=5; % Synaptic weight
pesoefa=2; % Ephaptic Exponent (0- ephaptic OFF; 2 - ephaptic ON)


% Numeric Constants
dt = 0.001; % step (dt)
tp = 0:dt:t_final; % time
N_steps = length(tp); % Steps Number


%% Simulation
for zz = simul
    tic
    for pr = 1:length(prob)
        [s,t] = C_sw(n,viz,prob(pr)); % Adjacency Matrix generated by the function (Synapses)
        ad1 = zeros(n); % adjacency matrix
        for i = 1:length(s(:,1))
            for o = 1:length(t(1,:))
                if s(i,1)~=t(i,o)
                    ad1(i,t(i,o)) = 1;
                end
            end
        end
        adj = ad1'+ad1; % Symmetric Matrix
        for syn = 1:1:length(peso)
            for y = 1:length(pesoefa)
                v = zeros(1,N_steps); % LFP Vector
                x = zeros(n,N_steps); % Neurons Time series
                %% defining communications: Topology area
                g = zeros(n,N_steps); % Count Synaptic Inputs
                fired = zeros(n,N_steps); % Spikes Graphic
                gamma = peso(syn); % Ephaptic Coupling Intensity.
                fire = zeros(n,N_steps); % Spikes Sum for each neuron
                
                % Model Constants
                a = linspace(23.75,27.25,n);
                b = linspace(28.5,31.5,n);
                c = C_swe(n); % Adjacency Matrix generated by the function (Ephaptic)
                if pesoefa(y) == 0
                    Amp = 0; % Ephaptic OFF
                else
                    Amp = 5; % Ephaptic ON
                end
                c = c.*Amp*10^(-1*pesoefa(y)); % Ephaptic weight
                I = zeros(n,N_steps); % Ephaptic Sum for each neuron
                % Defining the model and the variables
                f1 = @(tp,g,x,a,b,I) (a*x^2+b*x+9+g-I); % QIF-E equation
                
                % Initial Conditions
                x0 = zeros(1,n); % Initial Conditions for each neuron
                tic % Simulation Time Count (MATLAB FUNCTION)
                % For each neuron
                for i=1:N_steps-1
                    x(:,1) = x0(:); % Assigning Initial Conditions
                    g = (gamma*(mtimes(adj,fire))); % Synpatic Inputs Sum
                    %% Euler Integration
                    for k = 1:n
                        for h = 1:n
                            if k ~= h
                                I(k,i) = I(k,i)+(c(k,h)*(x(k,i)-x(h,i))); % Ephaptic Coupling Sum (Superposition Principle)
                            else
                                I(k,i) = I(k,i); % Ephaptic Coupling Sum
                            end
                        end
                        x(k,i+1) = x(k,i) + f1(tp(i),g(k,i),x(k,i),a(k),b(k),I(k,i))*dt; % Euler Integration
                        %% Reset Condition and Synaptic simulation
                        if x(k,i+1) >= 90
                            x(k,i) = 90; % Peak
                            x(k,i+1) = -5; % Hyperpolarization
                            for p = i:i+20
                                fire(k,p) = exp(-(p-i)/6); % Neighbor synaptic function
                            end
                            fired(k,i) = k;
                            
                        end
                    end
                end
                
                %%
                v = mean(x); % LFP
                %% Graphic Section
                figure1 = figure('Color',[1 1 1]);
                plot(tp,v,'-k','LineWidth',2)
                xlabel('Time(s)')
                ylabel('LFP(mV)')
                figure2 = figure('Color',[1 1 1]);
                for q =1:n
                    hold on
                    plot(tp,fired(q,:),'.k','MarkerSize',8)
                end
                xlabel('Time(s)')
                ylabel('# neuron')
                ylim([.2 n+.2])
                xlim([0 tp])
            end
            
        end
    end
end

toc


%% Small-World Topology (synaptic communication)
% Copyright 2015 The MathWorks, Inc.

function [s,t] = C_sw(N,K,beta)
% H = WattsStrogatz(N,K,beta) returns a Watts-Strogatz model graph with N
% nodes, N*K edges, mean node degree 2*K, and rewiring probability beta.
%
% beta = 0 is a ring lattice, and beta = 1 is a random graph.

% Connect each node to its K next and previous neighbors. This constructs
% indices for a ring lattice.
s = repelem((1:N)',1,K);
t = s + repmat(1:K,N,1);
t = mod(t-1,N)+1;

% Rewire the target node of each edge with probability beta
for source=1:N
    switchEdge = rand(K, 1) < beta;
    
    newTargets = rand(N, 1);
    newTargets(source) = 0;
    newTargets(s(t==source)) = 0;
    newTargets(t(source, ~switchEdge)) = 0;
    
    [~, ind] = sort(newTargets, 'descend');
    t(source, switchEdge) = ind(1:nnz(switchEdge));
end


end
%% All-to-all weigthed Topology (Ephaptic Coupling)
function [adje] = C_swe(n)

aux = zeros(n);

for j = 1:(n+2)/2
    if 1 ~= j
        aux(1,j)=(1/abs(1-j));
    end
end
for k = n:-1:(n+2)/2
    if 1 ~= k
        aux(1,k) = aux(1,abs(n-k+2));
    end
end


for i = 2:n
    for l = i:n
        aux(i,l)=aux(i-1,l-1);
    end
end

aux = aux'+aux;
adje=aux;
end

\end{lstlisting}

%\section*{Acknowledgments}
%GMC is recipient of the Conselho Nacional de Desenvolvimento Científico e Tecnológico (CNPq - Brazil) fellowship (\#140895/2021-3). 
%GC is recipient of the Conselho Nacional de Desenvolvimento Científico e Tecnológico (CNPq - Brazil) fellowship (\#307907/2019-8). 
%MPBS is recipient of the Coordenação de Aperfeiçoamento de Pessoal de Nível Superior (CAPES - Brazil) fellowship (\#88887.900715/2023-00).
%GZDL is recipient of the Conselho Nacional de Desenvolvimento Científico e Tecnológico (CNPq - Brazil) fellowship (\#309440/2022-0).

%\nolinenumbers

% 

\end{document}